\documentclass{article}

\usepackage{arxiv}

\usepackage[utf8]{inputenc} 
\usepackage[T1]{fontenc}    
\usepackage{hyperref}       
\usepackage{url}            
\usepackage{booktabs}       
\usepackage{amsfonts}       
\usepackage{nicefrac}       
\usepackage{microtype}      
\usepackage{lipsum}		    
\usepackage{hyphenat} 
\usepackage{graphicx}
\usepackage{natbib}
\usepackage{doi}

\newcommand{\pv}[1]{\texttt{#1}}      
\newcommand{\prop}[1]{\texttt{#1}}    

\title{A Core Ontology for Particle Accelerators: Interoperable Data and Workflows Across Facilities}

\date{\vspace{-2em}November 21, 2025}

\author{ \href{https://orcid.org/0000-0003-3814-8417}{\includegraphics[scale=0.06]{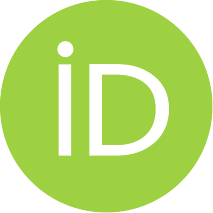}\hspace{1mm}Chris ~Tennant} \\
	Jefferson Laboratory\\
	12000 Jefferson Ave\\
	Newport News, VA 23606 \\
	\texttt{tennant@jlab.org} \\
}

\renewcommand{\headeright}{}
\renewcommand{\undertitle}{}

\hypersetup{
pdftitle={A Core Ontology for Particle Accelerators: Interoperable Data and Workflows Across Facilities},
pdfsubject={},
pdfauthor={Chris ~Tennant},
pdfkeywords={ontology, knowledge graph, particle accelerator, interoperability, data integration},
}

\begin{document}
\maketitle

\begin{abstract}
We propose a small, shared core ontology for particle accelerators that provides a semantic backbone for interoperable data and workflows across facilities. The ontology names key device types, signals, parameters, and regions, and relates them through explicit properties (e.g., \prop{hasSetpoint}, \prop{hasReadback}, \prop{partOf}). Each site contributes a lightweight facility bundle – a profile that maps local conventions into the shared vocabulary plus data slices that instantiate those mappings – without renaming channel addresses or changing existing systems. Using standard W3C technologies, the approach supports both sparse and rich descriptions. We demonstrate the idea on two beamline segments at different laboratories. A single semantic query is expressed once and evaluated against both knowledge bases, returning the locally correct PVs. The ontology thereby enables not only portable workflows but also interoperable data, since measurements and catalogs are annotated with shared semantics rather than facility-specific names. The framework complements, rather than replaces, existing middle layers and lattice/data standards, and it creates a stable foundation for reusable tools and agentic workflows.
\end{abstract}


\section{Introduction}
Accelerator information is abundant – device catalogs, channel names, and simulation outputs – but it is encoded in facility-specific ways that resist reuse. Names, suffix conventions, and schemas that make perfect sense locally often break when a script or workflow is moved across facilities. What’s missing is a shared, machine-readable description of what the objects are and how they relate, so tools operate on meaning rather than on local conventions. We propose a small core accelerator ontology as that semantic backbone. It lets each facility map its existing systems into common concepts (devices, signals, parameters, regions) without renaming anything. The same structure accommodates both sparse and rich descriptions. One facility may expose only setpoint channels for quadrupoles, while another may attach hardware details, survey coordinates, and expected optics values. 

To make the idea concrete, we show a minimal cross-facility example: two beamline segments, independently named and managed, are queried through the same semantic intent and return the locally correct results. The goal is not to standardize formats or nomenclature, but to demonstrate how a lightweight, shared ontology enables interoperable data and workflows across facilities.

\section{Key Definitions}
\label{sec:definitions}
In this context, a domain ontology for accelerators is a relatively small, curated vocabulary that achieves the following:

\begin{itemize}
    \item Names the key kinds of things we care about (quadrupoles, correctors, BPMs, ion pumps, power supplies, signals, etc.)
    \item States how those things are related (a \pv{BPM} \textit{is a kind of} \pv{Diagnostic}, \pv{IPM0R03.XPOS} \textit{measures} beam position, \pv{IPM0R03.XPOS} \textit{has units} of mm)
    \item Provides clear, machine-readable definitions so that software can interpret these concepts consistently
\end{itemize}

An ontology is more structured than a simple list of acronyms or channel names, and more general than a single database schema. It describes the types and relations in the domain, not individual instances or one particular storage layout.

Compared to a traditional relational database, we do not start with a fixed table layout. Instead, we begin by defining an ontology – a set of classes and relationships that captures how accelerator components, signals, and quantities are organized. From there we add triples as needed. This ontology naturally expresses hierarchies (for example, \pv{Quadrupole} is a \prop{subClassOf} \pv{Magnet}), part-of structures (this magnet belongs to that beamline), and cross-facility mappings (the local device family \prop{MQD*} corresponds to the core concept \pv{Quadrupole}). Once we attach facility-specific facts to this ontology, we obtain a knowledge graph that can evolve over time without schema migrations as new device types or relationships are added. Traditional databases and property-graph stores remain useful, the ontology-backed knowledge graph is an additional semantic view of the same reality.

In this note, an ontology refers to the shared schema (classes and relationships) and knowledge graph refers to the resulting graph once that schema is populated with facility data. In our case study, the knowledge graph is represented using the W3C Resource Description Framework (RDF) \cite{cyganiak2014rdf11}, where every fact is a subject–predicate–object triple.

It is helpful to distinguish between the terminology box (TBox), which is the ontology itself, from the assertion box (ABox) which represents the concrete facts (\pv{MQD0R03} is a \prop{subClassOf} \pv{Quadrupole}, \pv{MQD0R03} \prop{hasSetpoint} \pv{MQD0R03.BDL}). This separation is what allows a single core ontology (TBox) to be reused across multiple facilities, with each site contributing its own ABox data that aligns local objects to the shared vocabulary.

To avoid inventing yet another bespoke format, we leverage standard W3C technologies. This choice has critical implications:

\begin{itemize}
    \item Interoperability: Any tool that understands RDF (e.g., graph databases, visualization libraries, semantic reasoners) can immediately work with our accelerator data
    \item Longevity: W3C standards evolve slowly and maintain backward compatibility. An ontology defined today will remain queryable decades from now
    \item Tooling Ecosystem: We inherit a mature stack of validators, reasoners, query engines, and visualization tools without writing them ourselves
\end{itemize}

Examples of the specific technologies are:

\begin{itemize}
    \item RDF as the core data model: facts expressed as triples
    \item RDFS \cite{brickley2014rdfs} and OWL \cite{owl2overview2012} to define classes, subclasses, and constraints
    \item SHACL \cite{knublauch2017shacl} to validate that each facility’s data conforms to the intended structure before tools rely on it
    \item ELK \cite{kazakov2014elk}, HermiT \cite{glimm2014hermit}, or Pellet \cite{sirin2007pellet} as OWL reasoning engines: the ability to infer logical consequences from RDF data and schemata
    \item SPARQL \cite{harris2013sparql11} as the query language, playing a role analogous to SQL but tailored to graph-shaped data and semantic relationships
\end{itemize}

The emphasis in this note is not the syntax of these standards, but the guarantee they provide. We are building on an open, well-documented stack that many tools already understand, making our accelerator ontology more portable and sustainable over time.

\section{Motivation}
\label{sec:motivation}

This note argues that a small, shared core ontology for particle accelerators can provide a robust semantic backbone for our data and workflows. The focus is on explaining what an ontology is in this context, why it is useful, and how it can coexist with existing systems. We use a concrete example – a cross-facility query over two different beamline segments – to anchor the discussion, but the implementation details are deliberately kept light. A full technical treatment will appear in a future publication.

This is not a proposal to standardize every file format or to dictate new naming conventions. It is also not a full tutorial on the Semantic Web stack. Rather, the goal is to motivate a complementary layer that captures, in a machine-readable way, the same conceptual structure that operators and physicists already use informally when they talk about the machine. 

The discussion assumes a working familiarity with beamlines, channel addresses, and middle layers, but no prior exposure to ontologies (interested readers are directed to Refs. \cite{arp2015bfo} \cite{allemang2020semanticweb} \cite{keet2020intro} for background on ontologies and their development). Many readers will recognize themes from earlier attempts to define common lattice or data standards, or from maintaining their own facility-specific middle layers, yet no more than a basic sense of device types and querying an archive or database is required.

Over the last three decades there have been repeated community efforts to define common, code-independent standards for accelerator lattice and data description formats – ranging from early standard input and exchange formats through XML-based schemas \cite{carey1984standardinput} \cite{grote1998sxf} \cite{malitsky1998adef} \cite{malitsky2006formats} \cite{sagan2006aml} \cite{mitchell2025pals}. The approach here is intentionally different. We do not ask anyone to change local channel names or internal schemas. Instead, we introduce a shared semantic layer that can sit alongside these systems and describe, in common terms, what the local objects mean. The core ontology is meant to complement lattice standards, middle layers, and file-format agreements, not to replace them.

\section{Interoperability Challenges in Data and Workflows}
\label{sec:interop}

Every facility has developed its own pragmatic solutions for keeping track of devices and signals: mnemonic channel names, hierarchical naming trees, database keys, and lookup tables. A quadrupole setpoint may be referenced as \pv{MQD0R07.S} at one facility and \pv{XFEL.MAGNETS/MAGNET.ML/Q.1088.L3/CURRENT.SP} at another \cite{mayet_private}. Setting and readback channels may be distinguished by suffixes or punctuation.

These conventions work well locally, because everyone internal knows the idioms. The problem appears as soon as we try to port an analysis script from one facility to another, combine data from two machines in a single study, or build agentic systems that are expected to operate across different sites. The underlying physics concept “quadrupole current setpoint” is the same, but the machine-readable representation is not.

Previous attempts to define a single standard (for lattices, file formats, or naming) have taught us several lessons. First, social and organizational constraints matter. Changing names or formats can break existing tools and operational procedures. Even if the long-term gain is clear, the short-term cost is real. And second, facilities have different needs and histories. A format or scheme that fits one machine may be too rigid, too verbose, or too sparse for another.

These challenges do not mean that standardization is impossible, but they suggest that approaches which demand uniformity everywhere are likely to stall. The key is to separate semantic alignment (“we agree that this is a quadrupole setpoint”) from syntactic uniformity (“we all call it \pv{Q:CURRENT:SET}”).

For many workflows, what we really need is semantic interoperability, that is, the ability for tools and agents to understand that different local objects play the same conceptual role, even if they are named or stored differently. More succinctly, syntactic interoperability is about matching formats and field names, whereas semantic interoperability is about matching meanings and relationships.

The approach in this note focuses on the latter. We propose to capture a carefully chosen slice of accelerator semantics in an explicit, shared model, and to map each facility’s local systems to that model. Once this is in place, a tool can ask “give me all magnets and their setting channel addresses” in a single semantic language, and the underlying system takes care of retrieving the right process variables from the right facility-specific representations. 

When we talk about interoperable data, we do not mean that every facility stores numbers in the same file format or uses the same channel names. We mean that device catalogs, signal catalogs, and time-series measurements at different facilities are all annotated with the same ontology classes and relations. A quadrupole current setpoint in CEBAF and a quadrupole current setpoint in RHIC may live in different archives under very different channel names, but once they are tied to the shared concepts \pv{Quadrupole}, \prop{hasSetpoint}, \prop{inSegment}, etc., tools can find, merge, and interpret them in a uniform way. In that sense, the ontology makes the underlying data products themselves interoperable, not just the workflows that operate on them.

\section{A Core Accelerator Ontology}
\label{sec:onto}

The initial core ontology is intentionally modest in scope. It focuses on a common subset of accelerator components that is both widely shared and operationally important: beamline devices (magnets, diagnostics, vacuum elements) and their associated control signals (setpoints, readbacks). This gives us a small but useful backbone that can support real workflows and then evolve as needed. When we say the ontology is “modest” or “lightweight”, we refer to the core TBox – the shared vocabulary of device classes and relationships – not to the facility-specific data slices. A facility with thousands of devices will naturally have a correspondingly large ABox, but all sites share the same compact set of core concepts.

Furthermore, “modest” does not mean “thin.” Even with this small set of classes, the ontology can host a great deal of detail when a facility wants it. The same quadrupole instance that participates in optics matching can also carry hardware metadata (physical dimensions, serial number, manufacturer, date installed), engineering information (survey and alignment coordinates), and simulation or design information (expected Twiss parameters and other beam properties at that element). The core simply provides the hooks. Facilities are free to attach as much or as little information as they find useful.

We do not attempt, at this stage, to model every subsystem. The core should be facility-agnostic yet realistic: its concepts must correspond to things that exist and are used across multiple machines. It should be small, but not smaller than needed – every class and relation should earn its place by supporting concrete queries or mappings. Relationships such as “setpoint,” “readback,” or “part of this sector” are represented explicitly rather than embedded in naming conventions. The structure is extensible so facilities can add detail without breaking the shared layer.

\begin{figure}
    \centering
    \includegraphics[width=1\linewidth]{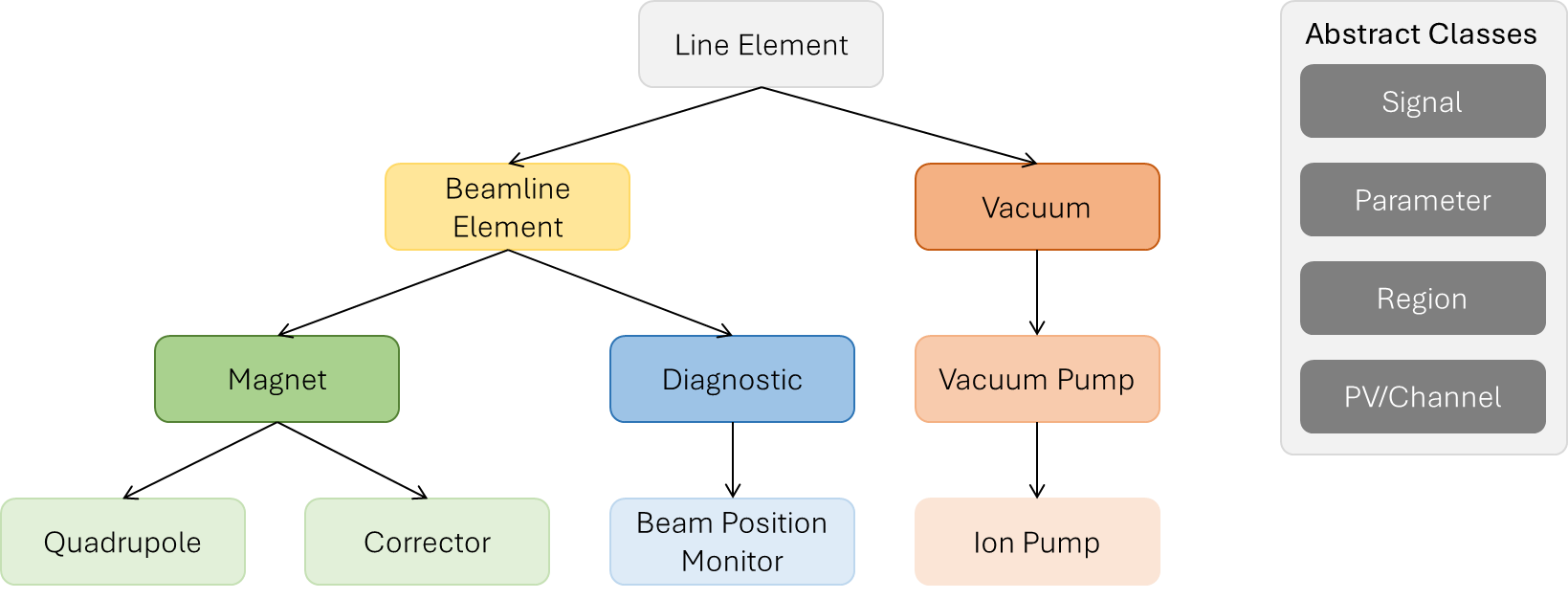}
    \caption{Hierarchical class structure showing \prop{subClassOf} relationships (denoted by arrows) in an accelerator core ontology.}
    \label{fig:fig1}
\end{figure}

At the heart of the core is a simple taxonomy. A representative slice is shown in Fig. \ref{fig:fig1}: Line Element at the top, with Beamline Element and Vacuum as major branches; Magnet and Diagnostic as two of the main subclasses of beamline element; and, beneath those, familiar concrete classes such as Quadrupole, Corrector, Beam Position Monitor, and Ion Pump. On the right-hand side, more abstract classes such as Signal, Parameter, Region, and PV/Channel capture the kinds of things we measure and control, independent of any one device family.

Device classes are linked to their signals and context. Setpoint and readback channels are associated via properties such as \prop{hasSetpoint} and \prop{hasReadback}. Devices are related to beamline regions or lattice elements through part-of relationships. This mirrors how we already describe the machine, but makes that structure explicit and machine-readable.

Throughout this note we use the EPICS term “process variable” (PV), which refers to a named control system channel that can be read or written over the network. A PV like \pv{MQD0R07.BDL} represents the desired magnetic field setpoint for a specific quadrupole. The full PV name, setpoint versus readback distinction, and unit conventions are all facility-specific – exactly the kind of detail the ontology abstracts away. Other control systems use equivalent concepts (TANGO attributes, DOOCS addresses), but we use “PV” as the generic term.

A key consequence is that the ontology must support both rich and sparse facility descriptions. Both descriptions are valid, and both participate in the same queries, with the richer instance simply contributing more facts for tools that know how to use them. The core ontology does not force a particular level of verbosity. In other words, the shared model defines what can be said, not what must be said. 

\section{A Facility Bundle}
\label{sec:bundle}

To make the core ontology useful in practice, each facility contributes a facility bundle:

\begin{enumerate}
    \item The shared core ontology (reused verbatim)
    \item A facility profile that maps local device types and PV naming patterns onto the core concepts
    \item One or more facility data slices that instantiate the profile with actual devices and channels for selected parts of the machine
\end{enumerate}

\begin{figure}
    \centering
    \includegraphics[width=1\linewidth]{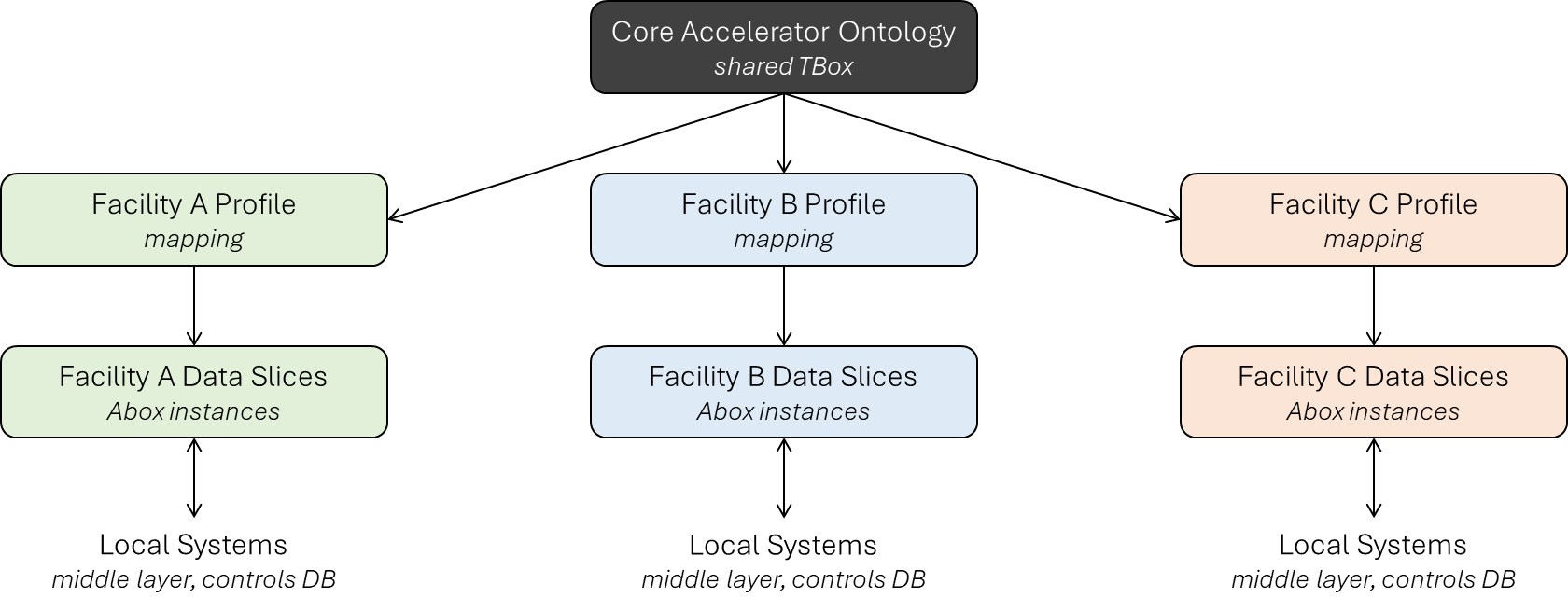}
    \caption{Schematic view of a shared core accelerator ontology and facility bundles. Each facility reuses the same core TBox, contributes its own profile (mapping local conventions into the shared vocabulary), and populates facility-specific data slices (ABox instances) derived from existing local systems. The core ontology is reused verbatim, only the profile and data differ.}
    \label{fig:fig2}
\end{figure}

Together, these components form a self-contained view of the facility that “speaks” the same semantic language as other bundles, even though the underlying controls and databases differ. See Fig. \ref{fig:fig2}.

\subsection{Facility Profiles}
A facility profile is where we capture statements such as: “Device family \pv{MQD*} is a kind of \pv{Quadrupole}” or “PVs matching pattern \pv{*.XPOS} measure horizontal position”. These mappings can be expressed as simple RDF/OWL axioms and SHACL rules. They can be relatively compact but are powerful. Once in place, they allow generic queries against the core ontology to be instantiated with local detail automatically. Different facilities can provide profiles of different richness. A minimal profile might only map magnets and BPMs, while a richer profile could also include RF, vacuum, and protection systems.

\subsection{Facility Data Slices}
The ABox portion of the bundle consists of data slices that populate the profile with concrete instances:

\begin{itemize}
    \item A device catalog: one row/node for each physical magnet, diagnostic, vacuum component, etc.
    \item A signal catalog: one row/node for each setpoint and readback PV that we care about
\end{itemize}

In the case study, such slices were built for beamline segments from CEBAF and the Advanced Light Source (ALS) by starting from existing inventories and middle-layer-like exports and converting them into RDF graphs. Practically, this amounted to writing a small number of translation scripts and then feeding the same pipeline with additional rows as more devices were included. The work is incremental, once the mapping from device-catalog entries to ontology classes and properties is established, extending coverage to additional sections of the machine is largely a matter of pointing the scripts at new devices.

The resulting graphs are loaded into standards-compliant triplestores and exposed as SPARQL endpoints, providing a uniform way to ask questions like: “List all quadrupoles in this beamline and their setting PVs” or “Find all BPM PVs that measure horizontal position.” Other interfaces – REST services, Python libraries, graphical browsers – can be layered on top. The critical point is that these tools talk to a shared semantic model, not directly to a site-specific naming scheme.

\section{Relationship to Traditional Middle Layers}
\label{sec:ml}

Middle layers – such as the Matlab Middle Layer (MML), PAMILA, and the more recent CATAP package – have been essential to operations and accelerator physics for decades \cite{portmann2005mml} \cite{liuzzo2025pythonml} \cite{hidaka2025pamilya} \cite{king2025catap}. They typically provide high-level functions, convenience wrappers for reading and writing PVs, and local data models that encapsulate naming rules, lattice information, and common device groupings. They are often the most concrete expression of how a machine is organized from the perspective of users and operational scripts.

Middle layers excel at local operations but face fundamental limitations for cross-facility interoperability:

\begin{itemize}
    \item Implementation Lock-in: Middle layers are typically language-specific (MML in MATLAB, PAMILA in Python) and tightly coupled to their runtime environments. The MML API at one facility cannot be directly consumed by tools at another facility without adaptation.
    \item Implicit Semantics: While a middle layer may group devices into families like \pv{Quadrupole} or provide methods like \prop{getSetpoint()}, these semantics exist only within that codebase. There is no machine-readable declaration that can be leveraged by external tools, agents, or other facilities.
    \item Point Solutions vs. Shared Infrastructure: Each facility develops its own middle layer independently. Even when using the same framework, the internal device models diverge. Cross-facility data integration requires custom mapping code for each pair of facilities.
    \item Limited Evolvability: Adding new device types or relationships in a middle layer often requires code changes across multiple modules. An ontology, by contrast, allows incremental extension through simple triple assertions.
\end{itemize}

These limitations are echoed in the middle-layer literature. Portmann’s description of MML, for example, shows how much semantic knowledge is concentrated in the Accelerator Object and Accelerator Data structures, yet tied to MATLAB and facility-specific configurations \cite{portmann2005mml}. Liuzzo and collaborators report similar challenges when exploring a shared Python middle layer across multiple labs \cite{liuzzo2025pythonml}, and King et al. emphasize both the benefits and constraints of CATAP’s YAML-based abstraction \cite{king2025catap}. In all cases, the hard work of “understanding the machine” lives inside code and configuration that is not easily shared or queried across facilities.

In many ways, mature middle layers already embody an implicit ontology. They recognize device types, distinguish between roles, and understand hierarchical groupings. The facility bundle concept deliberately overlaps with this conceptual structure. It aims to capture the same picture of “what exists and how it is organized,” but in a way that is explicit, declarative, and shareable beyond any single codebase or programming language.

However, there are important differences. A middle layer is usually an API and codebase, specific to a given language, runtime, and facility. Its internal model is rarely exposed in a machine-readable way that other tools – or other laboratories – can consume. An ontology plus knowledge graph, on the other hand, is a language-neutral, standards-based representation of semantics. It is not tied to a particular programming environment, and it is designed to be shared. This distinction is what makes an ontology more robust for cross-facility work. A core ontology, once defined, can be reused as-is, with each facility providing only its own profile and data slices. In short, middle layers are excellent for local scripting, while ontologies are better suited for interoperable semantics that must survive across time, tools, and institutions. 

A natural division of labor emerges: the ontology provides a canonical, machine-readable picture of “what exists, how things relate, and what roles channels play,” while middle layers consume this semantic foundation to offer higher-level operations (orbit correction, energy ramps, measurement sequences) and integrate with control-system APIs. Concretely, a middle layer at Facility A can be configured to read the facility bundle (profile + data slice) and automatically populate its internal device catalog. When a colleague from Facility B arrives with a script that queries “quadrupoles in the linac,” the ontology ensures both sites interpret that request identically, even though Facility A uses a MATLAB middle layer and Facility B uses CATAP. The middle layers handle the runtime operations and the ontology ensures semantic alignment.

As facilities evolve – magnets are replaced, new diagnostics installed, control system PVs renamed – only the facility data slice requires updating. The core ontology and profile remain stable. A new BPM can be added by asserting a handful of triples, no code changes are required in shared tools that query “all BPMs”. This contrasts with middle-layer-only approaches, where device additions often require modifying class hierarchies, configuration files, and accessor methods across multiple code modules.

\section{Demonstration: A Semantic Bridge Between Two Beamlines}
\label{sec:demo}

To test these ideas, we applied the core ontology to two modest but realistic testbeds, a simple beamline segment from CEBAF and the ALS. This analysis is used as part of a companion manuscript, currently in preparation ~\cite{apl_manuscript}.

For each, we started from existing inventories or middle-layer-like exports and constructed a facility profile and data slice aligned with the core ontology. The resulting graphs were loaded into triplestores and exposed as SPARQL endpoints, forming two separate but semantically compatible knowledge bases. 

The key step in each case was to map local device families and PV patterns to core concepts:

\begin{itemize}
    \item Local quadrupole families were asserted to be instances of the core class \pv{Quadrupole}
    \item Corrector families were mapped to \pv{Corrector}
    \item BPM records were mapped to \pv{BeamPositionMonitor}
    \item PVs with certain suffixes or metadata were identified as setpoints or readbacks and linked via \prop{hasSetpoint} and \prop{hasReadback} properties
\end{itemize}

Once these mappings were defined, the facility-specific differences in naming no longer mattered at the ontology level, both graphs spoke the same semantic language. Consider a simple request:

\prop{"List the setting PVs for all magnets in this beamline.”}

At the ontology level, the intent can be expressed as: (1) find all devices that are instances of \pv{Magnet} or its subclasses, and (2) for each, return the associated setting PV via the \prop{hasSetpoint} relation. This logic is the same for both facilities. The SPARQL queries differ only in the graph they target (CEBAF vs. ALS), not in their structure or meaning. Under the hood, the facility profiles ensure that the correct local PV names are returned, even though the channel naming schemes and device taxonomies differ. 

\begin{figure}
    \centering
    \includegraphics[width=1\linewidth]{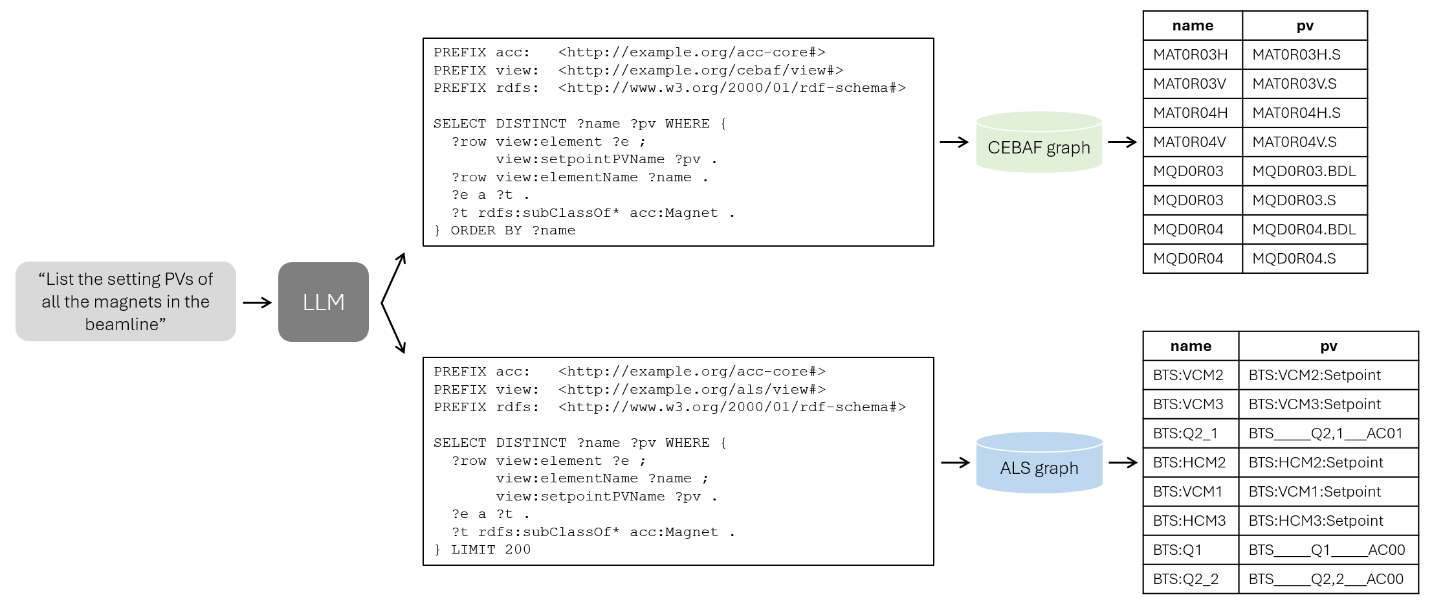}
    \caption{Example of an LLM transforming a natural language question into a formal SPARQL query. Despite the minor syntax differences, the two queries are functionally equivalent. When applied to each materialized graph, the correct PVs are returned for each facility.}
    \label{fig:fig3}
\end{figure}

On top of this, we added a thin natural-language layer. A local language model is given a concise description of the ontology and a few examples, and its sole job is to translate questions like the one above into SPARQL queries. This workflow is illustrated in Fig. \ref{fig:fig3}. Note that the LLM serves only as a convenience layer for translating natural language into SPARQL. The semantic query itself, finding devices of type Magnet with hasSetpoint relations, is facility-agnostic and resides entirely in the ontology. The LLM is simply reformatting the question into formal query syntax.

To illustrate the necessity of explicit semantics, consider what happens without an ontology:

\begin{itemize}
    \item A script written using Facility A’s middle layer requests “all quadrupole setpoints.” At Facility B, the device family is called “QUAD” not “Quadrupole.” The script fails immediately.
    \item An agentic system attempts to find BPMs that measure horizontal position. Facility A stores this in a PV called \pv{*.XPOS}, Facility B uses \pv{*.X\_RB}, and Facility C uses a TANGO attribute \pv{position/x}. Without semantic annotation, the agent cannot construct a cross-facility query.
    \item A machine learning model trained on archived data from multiple facilities attempts to correlate “quadrupole strength” with “beam position.” Without knowing that Facility A records strength in Tesla and Facility B in Amperes, and without knowing which channels represent strength vs. other magnet parameters, the training data is polluted with incompatible units and semantics.
\end{itemize}

The ontology resolves all three cases by providing explicit, queryable mappings from local conventions to shared concepts.

The outcome of the prototype is intentionally modest but important. First, the same semantic query pattern successfully returns the correct setting PVs for both beamlines. Second, no renaming of local channels was required; only mappings in the facility profiles. And thirdly, the logic for interpreting the question resides in the shared ontology, not in facility-specific code.

The experiment is deliberately small in scope, yet even at this scale it reveals where the core ontology is sufficient and where it needs more nuance; which local naming patterns are easy to map and which require additional metadata, and how much facility effort is required to create a useful profile. These lessons will guide any broader effort, but they already demonstrate the central claim of this note, a lightweight core ontology can support non-trivial, cross-facility workflows without demanding changes to local naming or middle layer code.

\section{Broader Implications and Opportunities}
\label{sec:implications}

There is growing interest in making accelerator data more Findable, Accessible, Interoperable, and Reusable (FAIR) \cite{wilkinson2016fair}. A core ontology approach directly supports interoperability, since different facilities can agree on the meaning of devices and signals even when their implementations differ. It also supports reusability, because tools, queries, and agentic workflows built on top of the ontology can be ported to new machines with far less effort than rewriting everything from scratch. 

Once we can reliably answer queries across facilities (e.g., “Which channels correspond to these devices and roles?”), a range of higher-level capabilities become possible, such as:

\begin{itemize}
    \item Shared analysis notebooks and dashboards that work at multiple labs
    \item Automated agents that can retrieve relevant channels, fetch archive data, and perform standard operational procedures without being hand-tuned for each site \cite{hellert2025agentic}
    \item Safer, more auditable AI-assisted operations, because the semantics are explicit and inspectable rather than buried in model weights or ad-hoc code
\end{itemize}

In the prototype, a language model plays a narrowly defined role by translating operator intent into SPARQL queries over the knowledge graph, which then feeds other tools (archive readers, plotting scripts). This separation of concerns keeps the ontology as the stable source of meaning, while letting the agentic layer evolve. 

A shared core ontology raises governance questions: who decides when to add new classes, and how are breaking changes handled? We envision a lightweight community process similar to those used for lattice file standards, with versioned releases and deprecation periods. Because facilities consume the ontology via profiles, they can choose when to adopt new versions, avoiding the “forced march” problem of centralized standards.

Many current initiatives – whether under the banner of agentic workflows, digital twins, or data-sharing projects – face the same basic hurdle: different facilities use different internal representations, and tool builders are forced to re-implement similar logic multiple times. By providing a shared semantic backbone that is small and realistic, grounded in widely used W3C standards, and designed to respect local autonomy, a core accelerator ontology can serve as a common thread through these efforts.

\section{Summary}
\label{sec:summary}

We have made a conceptual case for introducing a core ontology for particle accelerators, supported by facility-specific profiles and data slices. Our current landscape of naming conventions, lookup tables, and middle layers solves many local problems, but it also makes cross-facility interoperability difficult. A domain ontology addresses this by providing a compact, explicit description of key device types, signals, and relationships, expressed using open standards. When each facility maps its existing systems into this shared semantic model, we can pose meaningful questions and answer them with the same query pattern across different machines.

Ontologies and middle layers are not competing technologies but complementary ones. Middle layers remain indispensable for local scripting and control, while the ontology offers a more robust, facility-agnostic foundation for data integration, shared tools, and agentic workflows. The prototype with two beamlines is small but concrete, and it demonstrates that the approach is feasible without disruptive changes to existing infrastructure.

\section*{Acknowledgments}
A special thanks to Thorsten Hellert (LBNL), Antonin Sulc (LBNL), and Frank Mayet (DESY) for many useful conversations. This work was supported by the U.S. Department of Energy, Office of Science, Office of Nuclear Physics under Contract No. DE-AC05-06OR23177.

\bibliographystyle{unsrtnat}
\bibliography{references}  

@misc{cyganiak2014rdf11,
  author       = {Richard Cyganiak and David Wood and Markus Lanthaler},
  title        = {RDF 1.1 Concepts and Abstract Syntax},
  institution  = {W3C},
  year         = {2014},
  howpublished = {\url{https://www.w3.org/TR/rdf11-concepts/}}
}

@misc{brickley2014rdfs,
  author       = {Dan Brickley and R. V. Guha},
  title        = {RDF Schema 1.1},
  institution  = {W3C},
  year         = {2014},
  howpublished = {\url{https://www.w3.org/TR/rdf-schema/}}
}

@misc{owl2overview2012,
  author       = {{W3C OWL Working Group}},
  title        = {{OWL} 2 Web Ontology Language Document Overview (Second Edition)},
  institution  = {W3C},
  year         = {2012},
  howpublished = {\url{https://www.w3.org/TR/owl2-overview/}}
}

@misc{knublauch2017shacl,
  author       = {Holger Knublauch and Dimitris Kontokostas},
  title        = {Shapes Constraint Language ({SHACL})},
  institution  = {W3C},
  year         = {2017},
  howpublished = {\url{https://www.w3.org/TR/shacl/}}
}

@article{kazakov2014elk,
  author  = {Yevgeny Kazakov and Markus Kr{\"o}tzsch and Franti{\v{s}}ek Siman{\v{c}}{\'\i}k},
  title   = {The Incredible {ELK}: From Polynomial Procedures to Efficient Reasoning with {EL} Ontologies},
  journal = {Journal of Automated Reasoning},
  volume  = {53},
  number  = {1},
  pages   = {1--61},
  year    = {2014},
  doi     = {10.1007/s10817-013-9296-3}
}

@article{glimm2014hermit,
  author  = {Birte Glimm and others},
  title   = {HermiT: An {OWL} 2 Reasoner},
  journal = {Journal of Automated Reasoning},
  volume  = {53},
  number  = {3},
  pages   = {245--269},
  year    = {2014},
  doi     = {10.1007/s10817-014-9305-1}
}

@article{sirin2007pellet,
  author  = {Evren Sirin and others},
  title   = {Pellet: A Practical {OWL}-{DL} Reasoner},
  journal = {Web Semantics: Science, Services and Agents on the World Wide Web},
  volume  = {5},
  number  = {2},
  pages   = {51--53},
  year    = {2007},
  doi     = {10.1016/j.websem.2007.03.004}
}

@misc{harris2013sparql11,
  author       = {Steve Harris and Andy Seaborne},
  title        = {{SPARQL} 1.1 Query Language},
  institution  = {W3C},
  year         = {2013},
  howpublished = {\url{https://www.w3.org/TR/sparql11-query/}}
}

@book{arp2015bfo,
  author    = {Robert Arp and Barry Smith and Andrew Spear},
  title     = {Building Ontologies with Basic Formal Ontology},
  publisher = {MIT Press},
  year      = {2015}
}

@book{allemang2020semanticweb,
  author    = {Dean Allemang and James Hendler and Fabien Gandon},
  title     = {Semantic Web for the Working Ontologist: Effective Modeling for Linked Data, {RDFS}, and {OWL}},
  edition   = {3},
  publisher = {Association for Computing Machinery},
  address   = {New York, NY, USA},
  year      = {2020}
}

@book{keet2020intro,
  author    = {C. Maria Keet},
  title     = {An Introduction to Ontology Engineering},
  year      = {2020},
  note      = {Open textbook}
}

@inproceedings{carey1984standardinput,
  author    = {D. C. Carey and F. C. Iselin},
  title     = {A standard input language for particle beam and accelerator computer programs},
  booktitle = {Proceedings of the 1984 Summer Study on the Design and Utilization of the Superconducting Super Collider},
  address   = {Snowmass, CO, USA},
  pages     = {389--391},
  year      = {1984}
}

@techreport{grote1998sxf,
  author      = {H. Grote and others},
  title       = {{SXF} (Standard eXchange Format): definition, syntax, examples},
  institution = {Brookhaven National Laboratory},
  number      = {RHIC/AP/155},
  address     = {Upton, NY, USA},
  year        = {1998}
}

@inproceedings{malitsky1998adef,
  author    = {Nikolay Malitsky and Richard Talman},
  title     = {Accelerator Description Exchange Format},
  booktitle = {Proceedings of the International Computational Accelerator Physics Conference (ICAP'98)},
  address   = {Monterey, CA, USA},
  year      = {1998}
}

@inproceedings{malitsky2006formats,
  author    = {Nikolay Malitsky and Richard Talman},
  title     = {Accelerator description formats},
  booktitle = {Proceedings of the International Computational Accelerator Physics Conference (ICAP 2006)},
  address   = {Chamonix, France},
  year      = {2006}
}

@inproceedings{sagan2006aml,
  author    = {David Sagan and others},
  title     = {The Accelerator Markup Language and the Universal Accelerator Parser},
  booktitle = {Proceedings of EPAC 2006},
  address   = {Edinburgh, Scotland},
  pages     = {2278--2280},
  year      = {2006}
}

@inproceedings{mitchell2025pals,
  author    = {C. Mitchell and others},
  title     = {A community effort toward a Particle Accelerator Lattice Standard (PALS)},
  booktitle = {Proceedings of NAPAC2025: North American Particle Accelerator Conference},
  address   = {Sacramento, CA, USA},
  pages     = {350--353},
  year      = {2025},
  doi       = {10.18429/JACoW-NAPAC2025-TUP004}
}

@unpublished{mayet_private,
  author = {Frank Mayet},
  title  = {Private communication},
  note   = {Private communication},
  year   = {2025}
}

@inproceedings{portmann2005mml,
  author    = {G{\"u}nter Portmann and John Corbett and Andrei Terebilo},
  title     = {An Accelerator Control Middle Layer Using {MATLAB}},
  booktitle = {Proceedings of the 2005 Particle Accelerator Conference (PAC'05)},
  address   = {Knoxville, Tennessee},
  pages     = {4009--4011},
  year      = {2005},
  doi       = {10.1109/PAC.2005.1591699}
}

@inproceedings{liuzzo2025pythonml,
  author    = {Simone Liuzzo and others},
  title     = {Exploratory Tests for the Design of a Python Accelerator Middle Layer},
  booktitle = {Proceedings of the 16th International Particle Accelerator Conference (IPAC'25)},
  address   = {Taipei, Taiwan},
  publisher = {JACoW Publishing},
  pages     = {224--227},
  year      = {2025},
  doi       = {10.18429/JACoW-IPAC25-MOPB088}
}

@inproceedings{hidaka2025pamilya,
  author    = {Y. Hidaka and D. Allan and M. Rakitin},
  title     = {A New Python Middle Layer Framework: Particle Accelerator MIddle LAyer (PAMILA)},
  booktitle = {Proceedings of the North American Particle Accelerator Conference (NAPAC2025)},
  address   = {Sacramento, CA, USA},
  year      = {2025},
  doi       = {10.18429/JACoW-NAPAC2025-MOP005}
}

@article{king2025catap,
  author  = {M. King and others},
  title   = {Controls Abstraction Towards Accelerator Physics: A Middle Layer Python Package for Particle Accelerator Control},
  journal = {arXiv preprint},
  eprint  = {2509.19794},
  archivePrefix = {arXiv},
  year    = {2025}
}

@article{wilkinson2016fair,
  author  = {Mark Wilkinson and others},
  title   = {The {FAIR} Guiding Principles for Scientific Data Management and Stewardship},
  journal = {Scientific Data},
  volume  = {3},
  pages   = {160018},
  year    = {2016},
  doi     = {10.1038/sdata.2016.18}
}

@article{hellert2025agentic,
  author        = {Thorsten Hellert and others},
  title         = {Agentic {AI} for Multi-Stage Physics Experiments at a Large-Scale User Facility Particle Accelerator},
  journal       = {arXiv preprint},
  eprint        = {2509.17255},
  archivePrefix = {arXiv},
  year          = {2025},
  doi           = {10.48550/arXiv.2509.17255}
}

@unpublished{apl_manuscript,
  author = {Thorsten Hellert and others},
  title  = {Semantic Channel Finding: Bridging the Gap Between Control Systems and Natural Language},
  year   = {2025},
  note   = {Manuscript in preparation}
}

\clearpage
\section*{Glossary}

\textbf{ABox (Assertion Box)}: The set of concrete facts (instances) that populate the ontology, e.g., specific devices, signals, and their links to PVs and regions for a facility or data slice. 

\textbf{Description Logic (DL)}: the family of logic-based knowledge-representation formalisms that underpins OWL 2 DL.

\textbf{Domain Ontology}: A curated vocabulary for the accelerator domain that names key classes (devices, signals, parameters, regions) and the relations between them so software can interpret meaning consistently. 

\textbf{Elk}: An OWL reasoner optimized for the OWL 2 EL profile; used to efficiently infer class memberships and subclass relations in large, lightweight ontologies. 

\textbf{HermiT}: A complete OWL 2 DL reasoner used to compute logical consequences – including complex class expressions – over OWL ontologies. 

\textbf{Knowledge Base}: The semantic environment that manages, interprets, and queries a knowledge graph. A knowledge base includes the knowledge graph as well as the accompanying reasoning services, inference rules, constraints, and the query and storage infrastructure.

\textbf{Knowledge Graph}: A structured representation of classes, properties, and instance data expressed as an RDF graph. It consists of the ontology (TBox) together with its instantiated entities and relations (ABox), serialized as RDF triples.

\textbf{Middle Layer}: Facility-specific software that provides high-level APIs for accelerator control, abstracting low-level control system details. Middle layers excel at local scripting but lack cross-facility semantic interoperability without an ontology foundation.

\textbf{Ontology}: A formal, machine-readable model that defines classes and relations.

\textbf{OWL (Web Ontology Language)}: The W3C language for defining ontologies (classes, properties, constraints).

\textbf{Pellet}: A practical OWL DL reasoner used to derive inferences from OWL ontologies. 

\textbf{PV (Process Variable)}: EPICS-specific name for a control system channel address. 

\textbf{RDF (Resource Description Framework)}: The W3C data model used to represent facts as subject–predicate–object triples. 

\textbf{RDFS (RDF Schema)}: Vocabulary for basic typing and subclass relations over RDF data, providing lightweight schema/typing used alongside OWL. 

\textbf{Reasoner}: Software that derives implicit facts from explicit RDF/OWL data (e.g., inferring that a device is a Magnet via subclass axioms or property restrictions). 

\textbf{Schema}: A structural description of data; in this note, either local database/middle-layer schemas or semantic schemas (RDFS/OWL) that define types and relations. 

\textbf{Semantic Interoperability}: The ability for tools to interpret different facilities’ data as the same things because they are annotated with shared ontology classes and relations. 

\textbf{Semantic Web}: The W3C standards stack (RDF, RDFS/OWL, SHACL, SPARQL, etc.) enabling machine-readable, linkable data and reasoning on the web. 

\textbf{SHACL (Shapes Constraint Language)}: A W3C language used to validate that facility data (ABox) conforms to the intended structure and constraints before tools rely on it. 

\textbf{SPARQL (SPARQL Protocol and RDF Query Language)}: The query language for RDF graphs, used to retrieve devices, signals, and relations from the knowledge bases (similar in role to SQL for relational data). 

\textbf{Taxonomy}: The hierarchical organization of classes (e.g., \prop{Magnet} $\supset$ \prop{Quadrupole}) connected by \prop{subClassOf} relations. 

\textbf{TBox (Terminology Box)}: The ontology itself (classes like \prop{Magnet}, \prop{BPM}; properties like \prop{hasSetpoint}, \prop{hasReadback}), separate from the ABox of instance data. 

\textbf{Triplestore}: A database optimized for RDF triples that stores the KB and exposes it via SPARQL endpoints. 

\textbf{W3C (World Wide Web Consortium)}: Standardizes the Semantic Web technologies (RDF, RDFS, OWL, SHACL, SPARQL) used in this work.






\end{document}